\documentclass[twocolumn,amsmath,aps,prl,preprintnumbers,amssymb,nofootinbib,showpacs,floatfix]{revtex4}
\usepackage{graphicx}
\usepackage{amsmath,amssymb,mathtools}
\usepackage{epstopdf}
\usepackage{subfigure}
\usepackage{color}
\usepackage{verbatim}
\graphicspath{{Write_up_figures/}}
\usepackage{ulem}

\DeclareMathOperator{\sinc}{sinc}
\def\bfl{{\boldsymbol \ell}}	
\def\bft{{\boldsymbol \theta}}	
\def\cpl{\chi}
\def\cpli{\chi_{i}}

\def\kpar{k_{\pl}}

\def\kperp{{\bf k}_{\perp}}

\def\rnui{r_{\nu i}} 
\def\tnu{\tilde{\nu}} 
\def\llangle{\left \langle}
\def\rrangle{\right \rangle}
\def\dt21{\delta T_{21}}
\def\dtone21{\delta T_{21}^{(1)}}
\def\dttwo21{\delta T_{21}^{(2)}}
\def\dttwob21{\overline {\delta T_{21}^{(2)}}}
\def\dtonecc21{\delta T_{21}^{(1)*}}
\def\dttwobcc21{\overline {\delta T_{21}^{(2)*}}}
\def\bTb{\bar T_b}

\def\bfk{{\bf k}}	
\def\bfq{{\bf q}}	
	
\def\bfr{{\bf r}}	
\def\dw21{{\delta^W_{21}}}
\def\d21{{\delta_{21}}}

\def\hone{H{\sc i} }
\def\honenosp{H{\sc i}}
\def\zoneHI{Z_{HI}}
\def\ztwoHI{Z^{(2)}_{HI}}
\def\boneHI{b^{(1)}_{HI}}
\def\btwoHI{b^{(2)}_{HI}}

\def\nn{\nonumber}
\def\kpar{{k_\parallel}}

\def\bfq{{\textbf{q}}}

\begin{document}
			
\title{A cross-bispectrum estimator for CMB-\hone intensity mapping correlations}
\author{Kavilan Moodley$^{1,2}$} \email{moodleyk41@ukzn.ac.za}
\author{Warren Naidoo$^{1}$}
\author{Heather Prince$^{1,3,4}$}
\author{Aurelie Penin$^{1}$}

\affiliation{$^1$ Astrophysics Research Centre \& School of
Mathematics, Statistics and Computer Science, University of KwaZulu-Natal, Durban, 4041, South Africa}
\affiliation{$^2$ Center for Computational Astrophysics, Flatiron Institute, 162 5th Avenue, 10010, New York, NY, USA}
\affiliation{$^3$ Department of Astrophysical Sciences, Peyton Hall, Princeton University, Princeton, NJ 08544, USA}
\affiliation{$^4$ Department of Physics and Astronomy, Rutgers, The State University of New Jersey, 136 Frelinghuysen Rd, Piscataway, NJ 08854, USA}

\date{\today}                                        

\begin{abstract}
Intensity mapping of 21cm emission from neutral hydrogen (\honenosp) promises to be a powerful probe of large-scale structure in the post-reionisation epoch. However, \hone intensity mapping (IM) experiments will suffer the loss of long-wavelength line-of-sight \hone modes in the foreground subtraction process. This significantly reduces \hone IM cross-correlations with projected large-scale structure tracers, such as CMB secondary anisotropies. 
Here we propose a cross-bispectrum estimator, $B^{\bar \kappa \dt21 \dt21},$ to recover the cross-correlation of the \hone IM field, $\dt21,$ with the CMB lensing field, $\kappa,$ constructed by correlating the position-dependent \hone power spectrum with the mean overdensity traced by CMB lensing. We study the cross-bispectrum estimator in the squeezed limit and forecast its detectability based on \hone IM measurements from HIRAX and CMB lensing measurements from Advanced ACT. We find that $B^{\bar \kappa \dt21 \dt21},$ in combination with the \hone IM and CMB lensing auto-spectra, can place sub-percent constraints on the growth rate of fluctuations, $f,$ and the small scale amplitude of fluctuations, $
\sigma_8.$ The cross-bispectrum, in combination with the auto-spectra and Planck priors, improves dark energy constraints to 0.025 on $w_0$ and 0.11 on $w_a$ for flat models. These results are robust to \hone foreground removal because they derive from small-scale \hone modes. The \honenosp-CMB lensing cross-bispectrum thus provides a novel way to recover \hone correlations with CMB lensing and constrain cosmological parameters at a level that is competitive with next-generation galaxy redshift surveys. As a striking example of this, we find a tight constraint of 27.8 meV (29.0 meV) on the sum of neutrino masses, while varying all redshift and standard cosmological parameters within a flat $\Lambda$CDM ($w_0w_a$CDM) model.
\end{abstract}

\maketitle

\section{Introduction}

Intensity mapping of the cosmic microwave background has provided exquisite measurements of linear cosmological modes projected along the line of sight, thereby enabling the most precise constraints on the cosmological model to date
\cite{spergel2003first, ade2016planck, das2011detection, ruhl2004south}. Going beyond these constraints will require probes of the three-dimensional large-scale structure that measure a much larger set of cosmological modes to high precision. Galaxy redshift surveys \cite{levi2013desi, spergel2015wide, laureijs2011euclid, adame2025desi} and post-reionisation hydrogen intensity mapping experiments \cite{battye2012bingo, bandura2014canadian, jonas2009meerkat, carilli2004science, chen2012tianlai, crichton2022hydrogen, vanderlinde2019lrp} targeting the baryon acoustic oscillation (BAO) signal as a probe of dark energy \cite{bull2015late,shaw2014all, santos2015cosmology}, will map the large-scale structure distribution out to high redshift and over large survey areas, thereby expanding our access to three-dimensional cosmological modes. 

Intensity mapping surveys of the 21cm hydrogen line \cite{bull2015late, camera2013cosmology} promise to be a relatively efficient probe for mapping large-scale structure using a single tracer over a large redshift range. However, large sky area HI intensity mapping experiments face unique challenges. The galactic synchrotron and extragalactic point source signals are several orders of magnitude larger than the cosmological HI signal \cite{santos2005multifrequency}. The proposed solution to removing these contaminants is to take advantage of their smooth power law frequency spectra by high-pass filtering the data in the frequency domain \cite{shaw2014all,liu2013global}, leaving behind the HI signal that is correlated in frequency over smaller separations, corresponding to the BAO scale along the line of sight. However, imperfect data calibration from a strongly chromatic interferometer threatens to leak power from smooth line-of-sight foreground modes into higher frequency \hone modes \cite{shaw2015coaxing, switzer2014erasing} so the focus in the field has primarily been on overcoming these systematic effects. 

These challenges have meant that the HI intensity mapping signal has not yet been detected in auto-correlation (though see \cite{paul2023first} for a recently reported detection on small-scales). However, we know the signal is present as it has been detected in cross-correlation with spectroscopic galaxy surveys \cite{chang2010hydrogen, masui2013measurement, rafiei2021chime, wolz2016intensity}. The astrophysical and cosmological constraints that could be provided by future HI cross-correlations have been studied in the literature \cite{pourtsidou2016prospects, pourtsidou2015cross,ansari2018inflation,shi2020hir4}, using the simplest two-point correlations of HI with either galaxy, cosmic shear or CMB lensing surveys. 

Cross-correlations of HI intensity mapping with CMB secondary anisotropies are interesting because of the unique physics, either gravitational or scattering, imprinted on the CMB by large-scale structure \cite{sarkar2009cmbr,tanaka2019detectability,sarkar2016redshifted,pourtsidou2017h}. However, due to the absence of large-scale line-of-sight modes in the HI signal as a result of foreground filtering, the cross-power spectrum between HI intensity mapping and the CMB is significantly reduced in the flat-sky geometry considered here. Recently, it has been proposed \cite{kothari2024wide, shen2025direct} that the two-point cross-correlation may be recovered by reverting to large-angle correlations, however, this is difficult for interferometers, such as HIRAX considered in this paper, which resolve out large angular scales. Recovering the two-point cross-correlation may be possible for single-dish experiments such as MeerKLASS \cite{Santos:2018JO} if the large-angle \hone signal can be adequately calibrated.

To recover the correlation between \hone intensity mapping and CMB secondary anisotropies, we propose the use of a higher-order correlation that takes advantage of the modulation of small-scale \hone modes by a large-scale density mode. An alternative, but related, approach is to reconstruct the long wavelength density modes using the small-scale \hone modes and then correlate this field with the projected CMB field \cite{zhu2016cosmic, karaccayli2019anatomy, zhu2022cosmic}. In this letter, we present a \honenosp-CMB lensing cross-bispectrum estimator that combines two HI fields, well-measured on small scales, and a CMB lensing convergence field. 

The rest of this letter is structured as follows. In section 1, we introduce the cosmic probes considered in this paper. In section 2, we present the HI-CMB lensing cross-bispectrum estimator. Finally, in section 3, we study cosmological parameter constraints from the HI-CMB lensing cross-bispectrum using the Fisher matrix. In our analysis, we assume the Planck 2018 cosmology and priors \cite{aghanim2020planck}: $h=0.67$, $\Omega_M = 0.315$, $\Omega_\Lambda = 0.684$, $\Omega_k = 0.0$, $n_s = 0.965$, $\sigma_8 =  0.811$, $w_0 = -1.03$ and $N_{eff} = 2.99.$ All distances and scales are expressed in physical (Mpc), rather than $h^{-1}$Mpc, units.

\section{Cosmic probes}

We consider the \hone intensity mapping signal and the CMB lensing convergence signal in a periodic comoving volume, $V_p(z_i)=\cpli^2  \, \rnui$, spanning a redshift slice centred at redshift $z_i,$ with width $\Delta z$ ($\sim 0.5$) corresponding to a dimensionless bandwidth, $\Delta \tilde \nu_i,$ and subtending a solid angle, $\Omega_i,$ on the sky. Working in this ``snapshot'' geometry \cite{smith2018ksz, bull2015late} we have $\cpli$ and $\rnui=\cpli/\tnu_i$ (where $\tnu_i=\nu_i/\nu_{21}$) defining the transverse and line-of-sight comoving distances, which project physical wavenumbers within the volume to angular and radial wavenumbers as $\kperp={\bfl / \cpli}$ and $\kpar={y / \rnui},$ respectively. 

The angular \hone signal in this volume is given by \cite{bull2015late}
\begin{equation}
\dt21(\bfl, y; z_i) = {\bTb(z_i)}
\zoneHI (\bfk; z_i) \, \delta_{m}(\bfk, z_i)/V_p(z_i),
\nonumber
\end{equation}
where $\bTb$ is the mean brightness temperature, $\zoneHI(\bfk; z_i) = \boneHI(z_i) + f(z_i) \mu_k ^2 $ includes the linear bias and redshift space distortion terms that relate the \hone density field to the underlying matter density field, $\delta_{m}$. Filtering the \hone signal to remove smooth-frequency galactic and extragalactic foregrounds \cite{santos2005multifrequency} removes low $k_\parallel$ modes in the \hone signal \cite{shaw2015coaxing, wang200621}, typically below a wavenumber $k_{\parallel} \sim 0.01$ Mpc$^{-1}.$ \cite{bull2015late}

The CMB lensing convergence, $\kappa (\bft) = \int d\chi \, \kappa (\bfr),$ is the projection of the matter density along the line-of-sight \cite{lewis2006weak}, where $\kappa (\bfr) = W_\kappa(\chi) \delta_m(\bfr)$ is given in terms of the lensing convergence redshift kernel, $W_\kappa (\cpl) = \frac{3}{2} \Omega_{m0} (H_0 \cpl / c)^2  (1+z)  \left(
\frac{\cpl_* - \cpl}{\cpl _*\cpl}\right),$ and $\cpl_*$ is the comoving distance to the last scattering surface. 
In harmonic space, the lensing convergence is given by 
\begin{equation}
\kappa (\bfl)  =    \int \frac{d\kpar}{(2\pi)} \int d \cpl \,e^{i  \kpar \cpl }\, K_\kappa(\cpl)  \, \frac{ \delta_{m}\left (\bfl/\cpl,\kpar, z=0 \right)}{\cpl^2}, \, 
\label{eq:convergence_harmonic}  \nonumber
\end{equation}
where $K_\kappa(\cpl) = D(\cpl) W_\kappa (\cpl)$ and $D$ is the growth function. 

We find that the cross-correlation between the projected CMB lensing signal, which has a broad redshift kernel supported by low $\kpar$ modes below $\sim 0.01$ Mpc$^{-1}$, and an HI intensity map, which has been cleaned of foregrounds and thus missing long-wavelength radial modes below $\sim 0.01$ Mpc$^{-1}$, is negligible due to the lack of overlap in large-scale radial modes. This is described further in Appendix A.

We can quantify this signal loss in terms of the signal-to-noise ratio for the cross-correlation of CMB and \hone IM experiments considered in this letter, specifically AdvACT \cite{Henderson_2016} 
and HIRAX \cite{newburgh2016hirax}, which will overlap over $\sim$ 15,000 deg$^2,$ and defer the study of future surveys to a follow-up paper \cite{naidoo2022}. The survey specifications and noise spectra for these experiments are presented in Appendix B, along with the signal power spectra and signal-to-noise ratio expressions. In Appendix B, we show that while the \hone intensity mapping power spectrum is detected with high significance, the cross-correlation spectrum detection significance falls drastically when low $\kpar$ \hone modes are removed. 

\begin{figure}[!t]
 % HI SNR
    \includegraphics[width=0.8\linewidth]{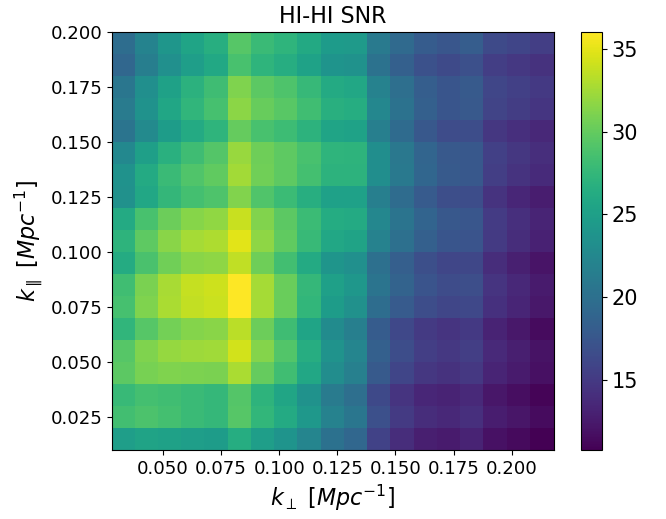}
 % Bispec SNR
	\includegraphics[width=0.8\linewidth]{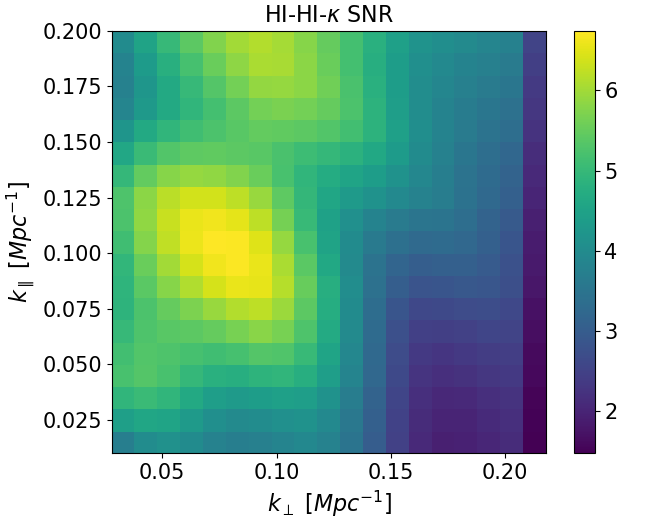} 
	\caption{Binned signal-to noise ratio (SNR) in the $k_\perp$-$k_\parallel$ plane for the $z=0.95$ redshift bin. 
	Top panel: SNR for the \hone power spectrum measured by HIRAX. Bottom panel: SNR for the cross-bispectrum measured by HIRAX and AdvACT. }
	\label{fig:HI Auto Bispec SNR 2D}
\end{figure}

\section{A cross-bispectrum estimator  to recover the cross-correlation of \hone intensity mapping with CMB lensing}

We can recover the long wavelength \hone modes required for the CMB lensing convergence cross-correlation by going to second order in the \hone field and taking advantage of the fact that short wavelength modes are correlated in the presence of a long wavelength mode \cite{bernardeau2002large,chiang2014position,chiang2015position}. This \textit{density modulation} effect results in small scale fluctuations being correlated with the large scale \hone background density, which is correlated with the CMB lensing convergence. In the large-scale structure literature, this effect has been well studied in terms of super-sample modes that contribute to super-sample variance in galaxy surveys (see \cite{takada2013power} and references therein).

We define the \honenosp-\honenosp-$\kappa$ cross-bispectrum as the correlation of the {\it projected} \hone position-dependent power spectrum with the average CMB convergence within that volume, 
\begin{equation}
\begin{aligned}
&B_{S,i}^{\bar \kappa \dt21 \dt21}(\ell, y; z_i) = \llangle  \left. P_{21}(\ell, y; z_i)  \right\vert_\bfr \bar{\kappa}(z_i) \vert_\bfr \rrangle,
\end{aligned}
\label{eqn:bispec}
\nonumber
\end{equation}
as detailed further in Appendix C, where the expectation value is taken over all sub-volumes across the survey region in a given redshift bin. The cross-bispectrum is dominated by squeezed configurations \cite{chiang2014position} for the large \hone wavenumbers of interest to us. In Appendix C we present the expression for the full cross-bispectrum, which is derived in   \cite{naidoo2022}, but focus here on its squeezed limit. We find that in the squeezed limit the \honenosp-\honenosp-$\kappa$ cross-bispectrum in redshift bin, $z_i,$ reduces to \cite{naidoo2022} 
\begin{equation}
\begin{aligned}
B_{S,i}^{\bar\kappa \dt21 \dt21}(\ell, y) &= V_B(\cpli)\, \mathcal{B}(k, \mu_k; f, b_1, b_2)\, P_{21}(\mathbf{k};z_i)  \\
 &\times \int q^2 {dq} W_{L,\kappa}(q) W_{L,21}(q) P_m(q;z_i) 
\end{aligned}
\nonumber
\end{equation}
where we have defined $V_B(\cpli) = {V_L W_\kappa(\cpli) / (V_p \cpli^2)} = \Omega_i \Delta \tilde \nu_i {W_\kappa(\cpli) / \cpli^2}$ and $\mathcal{B}(k, \mu_k; f, b_1, b_2)$ is given in Appendix C.
As noted in the literature \cite{chiang2014position, doux2016first}, the squeezed-limit bispectrum probes the linear response of the small-scale power spectrum, $P_{21},$ to the variance of the large-scale fluctuations, captured by the integral over $P_m$, with the response function for the \honenosp-\honenosp-$\kappa$ cross-bispectrum given by $\cal{B}$, and the overall normalisation set by appropriate projection and volume factors. 

In order to recover the cross-bispectrum with good significance we require short wavelength \hone modes that are well measured. This is the case for HIRAX, as seen in the top panel of Figure \ref{fig:HI Auto Bispec SNR 2D}, which shows that the \hone power spectrum modes in the range $ 0.05 \lesssim k/\text{Mpc}^{-1} \lesssim 0.15$ are measured with ${\rm SNR}\gtrsim 30$ in ($\kpar$, $k_\perp$) bins of width $0.01\,\text{Mpc}^{-1}.$ 

The \honenosp-\honenosp-$\kappa$ cross-bispectrum signal-to-noise ratio is discussed in Appendix D and plotted in Figure \ref{fig:HI Auto Bispec SNR 2D}. We see that the \honenosp-\honenosp-$\kappa$ cross-bispectrum is detectable with high significance, assuming HIRAX and AdvACT specifications, for a large range of radial and transverse modes. This is true even in the presence of the \hone foreground cut, unlike the cross-power spectrum. Furthermore, it is evident from Figure \ref{fig:HI Auto Bispec SNR 2D} that the cross-bispectrum estimator is fairly robust to the removal of foreground modes since the SNR is mainly contributed by scales outside the foreground wedge \cite{alonso2017calibrating}. Even when a conservative horizon-scale foreground wedge cut is applied we find that the total SNR in the $z=0.95$ bin only drops by about 10\% (from 88 to 80).

\section{Parameter forecasts}
\label{sec:forecasts}
We forecast constraints on parameters $p_a$ using the Fisher matrix in redshift bin $i$ given by 
\begin{equation}
\begin{aligned}
F_{ab,i} & = {1\over 2} S_{\rm area} \Delta \tilde{\nu}_i \int {d^2 \ell \over (2\pi)^2} \,   
\int {dy\over (2\pi)} \, {\partial_{p_a} {\cal S}_i(\ell,y)  \partial_{p_b} {\cal S}_i(\ell,y) \over {\cal V}_i(\ell,y)}\nonumber
\end{aligned}
\end{equation}
\cite{tegmark1997measuring,bull2015late}, where the signal spectra are ${\cal S}_i= \{ C_{S,i}^{\dt21}(\ell, y), C_S^{\kappa}(\ell), B_{S,i}^{\bar\kappa \dt21 \dt21}(\ell, y)\}$ and we use the diagonal covariances ${\cal V}_i(\ell,y)$ given in Appendix D. We combine constraints from the different power spectra and bispectra signals, and in different redshift bins, by directly adding the corresponding Fisher matrices, because the cross-probe covariances are negligible, as discussed in Appendix D. In this work we ignore the off-diagonal cross-covariance terms between the HI power spectrum and the bispectrum as we have shown that these contributions are neglegible \cite{naidoo2022}. The constraints in this section include a $k_{\parallel,cut}=0.01$ Mpc$^{-1}$ foreground cut for the HI signal. We include a range of parameters in our model but only discuss constraints for specific parameters of interest, as described next. 

\begin{figure}
	\centering
    \includegraphics[scale=0.31]{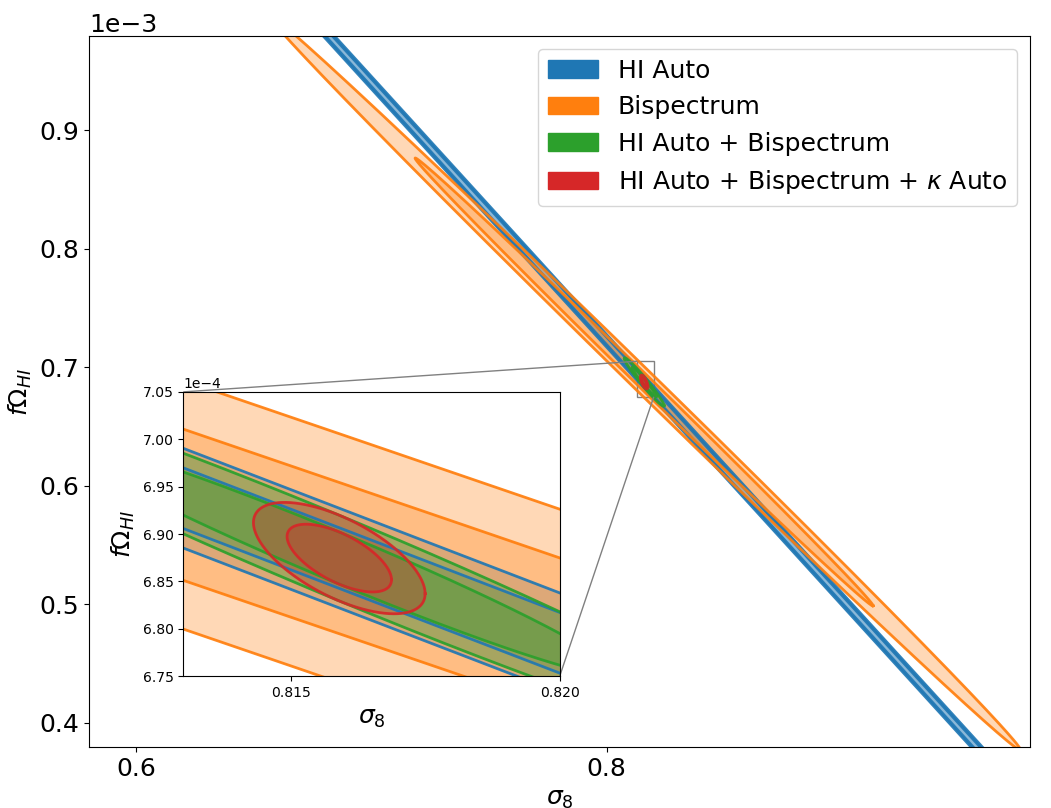}
    \includegraphics[scale=0.31]{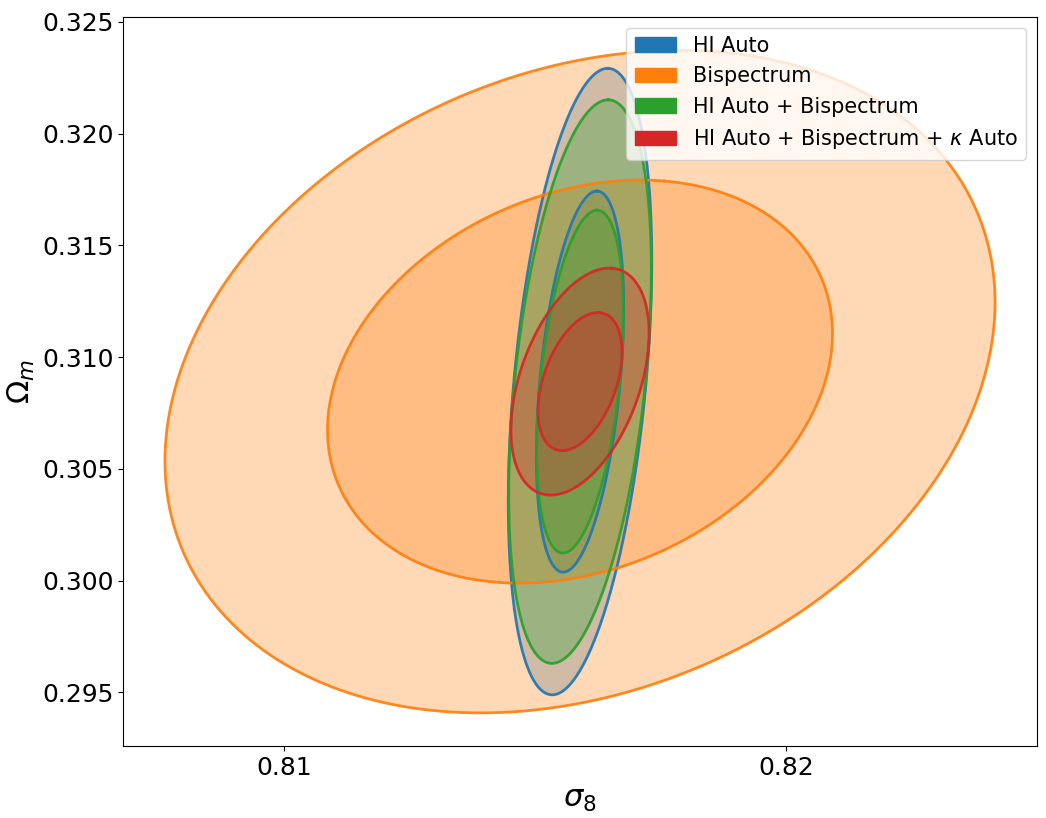}  
	\caption{Forecast 1$\sigma$ and 2$\sigma$ constraints on $f$ and $\sigma_8$ (top panel) from HIRAX \hone and AdvACT lensing, in the redshift bin centred at $z=0.95,$ for different power spectrum and cross-bispectrum combinations. This is one of four redshift bins spanning the HIRAX redshift range. The bispectrum has a different degeneracy direction to the \hone power spectrum in the $f$-$\sigma_8$ plane. The lensing power spectrum further constrains $\sigma_8$. We also show the $\Omega_m$ and $\sigma_8$ (bottom panel) constraints from  HIRAX \hone and AdvACT lensing for different power spectrum and cross-bispectrum combinations. }
	\label{fig:Constraints_f_sig8_omega}
\end{figure}

{$f$ and  $\sigma_8$}: We first vary the redshift dependent quantities $A_{\rm bao}, \sigma_8, f\Omega_{\text{\honenosp}}, b^{(1)}_{\text{\honenosp}}\Omega_{\text{\honenosp}}, b^{(2)}_{\text{\honenosp}}\Omega_\text{\honenosp}, d_A, \text{and } H,$ fixing their fiducial values using standard relations and giving them an independent value in each of the four redshift bins centred at $z=(0.81, 0.95, 1.27, 1.95).$ We focus on $f$ and  $\sigma_8$ constraints, marginalising over the other parameters. The combined spectra considered here are unable to constrain $\Omega_{\text{\honenosp}}$ independently of $f$, hence we consider the parameter $f\Omega_{\text{\honenosp}}.$ The top panel of Figure \ref{fig:Constraints_f_sig8_omega} shows the constraints on $f\Omega_{\text{\honenosp}}$ and $\sigma_8$ for the HI auto and bispectrum combined in the $z=0.95$ redshift bin, corresponding to 1\% and 0.4\% respectively. The cross-bispectrum helps to break the degeneracy between $f\Omega_{\text{\honenosp}}$ and $\sigma_8,$ because it has a fourth power dependence on $\sigma_8$ versus the squared dependence in the \hone power spectrum.  Adding CMB lensing reduces these to sub-percent constraints of (0.55\%, 0.34\%, 0.35\%, 0.50\%) on $f\Omega_{\text{\honenosp}}$ and (0.081\%, 0.080\%, 0.080\%, 0.081\%) on $\sigma_8,$ in the four redshift bins respectively, which significantly improve upon constraints from existing data \cite{de2017vimos,jullo2019testing, gil2016clustering} and are better than forecasted constraints from future surveys \cite{byun2020constraining}. 

{$\Omega_m$ and $\sigma_8$}: We now consider a vanilla $\Lambda$CDM model with parameters $\{\Omega_m, \sigma_8, h, n_s, \Omega_b\},$ where we marginalize over the HI biases and $A_{\rm bao}$, but retain constraints on $\{f, d_A, H,\sigma_8\}$ in each redshift bin. We introduce the distance scale parameters, $ \alpha_\perp \propto d_A^{-1}$ and
$ \alpha_\parallel \propto H $ \cite{blake2003probing}, and transform the Fisher matrix, $F_{ij},$ with the implicit redshift dependent functions and distance scale parameters, to the Fisher matrix, $F'_{ij},$ containing the cosmological parameters using
$ \left[ F'_{ij} \right] = \left[ M_{ij} \right]^T \left[ F_{ij} \right] \left[ M_{ij} \right] $ where  $M_{ij}={\partial_{p_i} / \partial_{p'_j}}.$
For all probes, we include \textit{Planck} priors from the Planck 2018 \cite{aghanim2020planck} temperature and polarization power spectra (but not lensing). The covariance between CMB lensing and the temperature and polarization constraints is negligible for the CMB surveys considered \cite{schmittfull2013joint}. We pre-marginalize over $\tau$ in the \textit{Planck} priors.  
The constraints on $\Omega_{m}$ and $\sigma_{8}$ are shown in the bottom panel of Figure \ref{fig:Constraints_f_sig8_omega} and in Table \ref{tab:LCDM_errors}. The $\Omega_m$--$\sigma_8$ figure of merit \cite{DESY3_2022} for this constraint is $\sim$ $10^5$, corresponding to a 0.32\% constraint on $S_8$, which will significantly improve on current large-scale structure and cosmic shear constraints, including with external data \cite{DESY3_2022, asgari2021kids, li2023hyper, heymans2021kids}. The improved constraint results from the cross-bispectrum  breaking the HI power spectrum degeneracy in these parameters, as can be seen in Figure \ref{fig:Constraints_f_sig8_omega}. This improvement enables tighter constraints on $h$ and $n_s,$ as evident in Table \ref{tab:LCDM_errors}.

\begin{table}[ht]
	\centering
	\begin{tabular}{|c|c|c|c|c|c|}
		\hline
		& $\Omega_m$ &$\sigma_8$& $h$ & $n_s$  &$\omega_b$\\
		\hline
	    \textit{Planck} & 0.0074 &0.0060&0.0054&0.0042&0.00015\\
		 \hline
		 \text{HIRAX \hone +} \textit{Planck}& 0.0057 & 0.00058 &0.0043& 0.0028 & 0.00014\\
		\hline
		 \text{Combined +} \textit{Planck}& 0.0021 & 0.00056 &   0.0022 & 0.0024 & 0.00013\\
		\hline
	\end{tabular}
	\caption{\label{tab:LCDM_errors} Marginalized 68\% cosmological parameter forecasts for the HI power spectrum and the combined constraint, that includes, in addition, the CMB lensing power spectrum and HI-CMB lensing cross-bispectrum, all with Planck priors.}
\end{table}

\begin{figure}[!t]
	\includegraphics[scale=0.28]{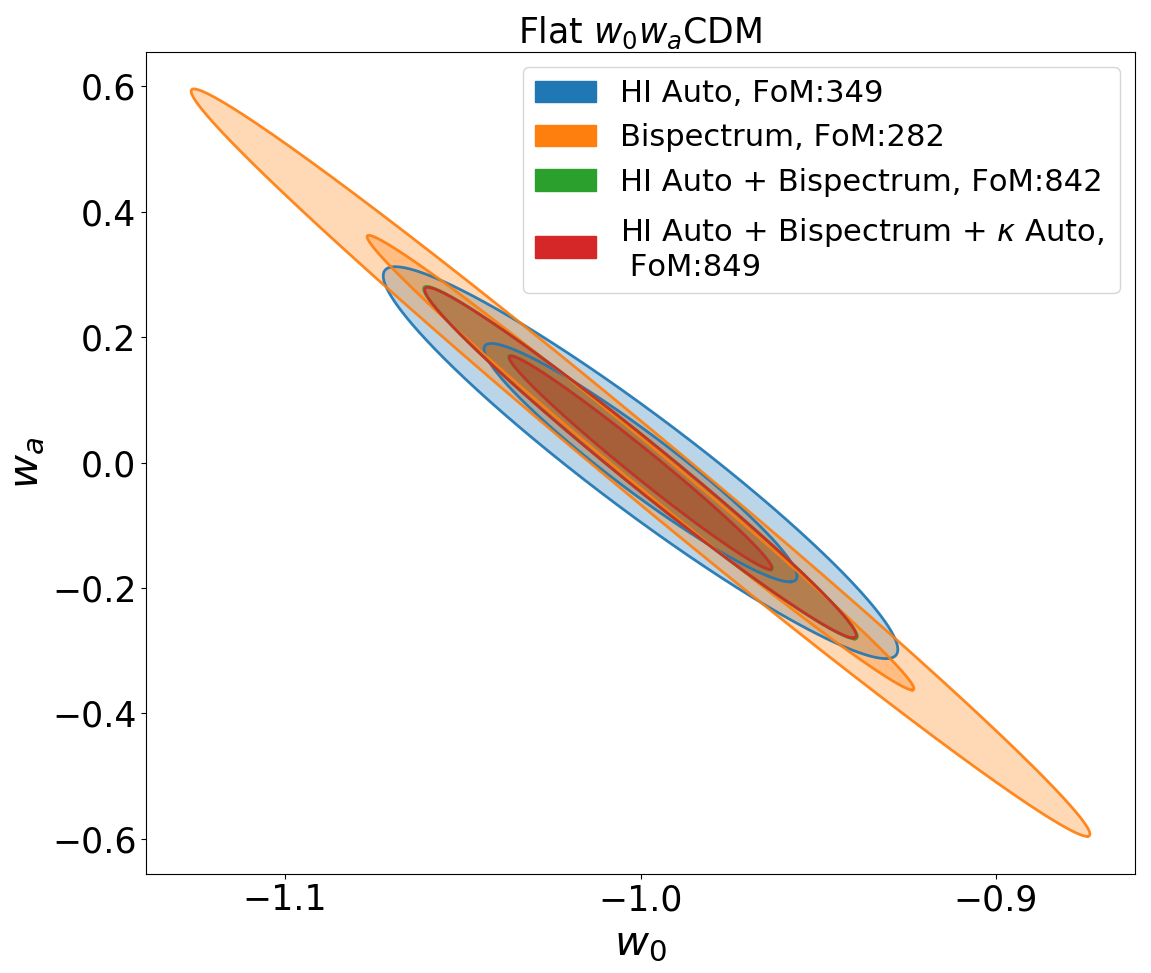} 
	\caption{Forecast 1$\sigma$ and 2$\sigma$ constraints on $w_0$ and $w_a$ from HIRAX \hone and AdvACT lensing for different power spectrum and cross-bispectrum combinations. 
    } 
	\label{fig:w0wa}
\end{figure}

{$w_0$ and $w_a$}: We next consider a dark energy equation of state, $w=w_0+w_a(1-a)$ \cite{chevallier2001accelerating,linder2003exploring}, in flat models with parameters $\{w_0, w_a, \Omega_m, \sigma_8, h\}.$
We present constraints on $w_0$ and $w_a,$ marginalising over the other parameters and imposing Planck 2018 priors \cite{aghanim2020planck} for all cases. The constraints presented include a $k_{\parallel}$ foreground cut and a wedge cut, showing that the cross-bispectrum constraints are indeed robust to HI foreground removal. As seen in Table \ref{tab:Cosmo_errors_wedge}, the cross-bispectrum combined with other probes is able to constrain $w_0$ and $w_a$ at the 2.5\% and 11\% level, respectively. These forecasts improve on recent BAO constraints in combination with CMB and SNe Ia data, which are $\sim 7\%$ and $\sim 50\%$ on $w_0$ and $w_a,$ respectively from the eBOSS cosmology analysis \cite{alam2021completed} and $\sim 7\%$ and $\sim 30\%$ respectively from the DESI cosmology analysis \cite{adame2025desi}. Our constraints are competitive with dark energy forecasts for SKA1-Mid \cite{bacon2020cosmology}, DESI \cite{aghamousa2016desi}, and Euclid \cite{ilic2022euclid}. The HI power spectrum and cross-bispectrum also break CMB lensing degeneracies, significantly improving CMB lensing constraints on all parameters listed in Table \ref{tab:Cosmo_errors_wedge}. As seen in Figure \ref{fig:w0wa}, adding the cross-bispectrum improves dark energy constraints due to the slightly different degeneracy direction relative to the HI power spectrum, which arises from the differing contributions of growth and distance scale parameters.  

\begin{table}[ht]
	\centering
	\begin{tabular}{|c|c|c|c|c|c|}
		\hline
		&	
        $\Omega_\Lambda$ & $w_0$ & $w_a$ &$h$\\
		 \hline
		 \text{HIRAX \hone} + \textit{Planck}
         & 0.0039 & 0.030 & 0.13 & 0.0026 \\
		\hline
         \text{AdvACT lensing} + \textit{Planck} 
         & 0.0043 & 0.59 & 2.15 & 0.0029 \\
        \hline
		 \text{Combined} + \textit{Planck}
         &  0.0023 & 0.025 &  0.11 & 0.0016\\
            \hline  
	\end{tabular}
	\caption{\label{tab:Cosmo_errors_wedge} Marginalized 68\% cosmological parameter forecasts for the HI power spectrum, the CMB lensing power spectrum and the combined constraint, that includes, in addition to these two power spectra, the HI-CMB lensing cross-bispectrum. All forecasts include Planck priors.} 
\end{table}

\section{Discussion} 

In this letter, we developed a new cross-bispectrum estimator that recovers correlations between HI intensity mapping surveys and projected cosmological fields like CMB secondary anisotropies. We applied this estimator to the cross-correlation of HI intensity mapping with CMB lensing, and found that it is detectable with high significance for the HIRAX and AdvACT surveys. The cross-bispectrum provides complementary information to the HI IM auto and CMB lensing power spectra that breaks degeneracies between key cosmological parameters. When combined with these probes, it provides cosmological constraints that are very competitive with future large-scale structure surveys, in particular on the growth function, the amplitude of small-scale fluctuations and  the dark energy equation of state. Moreover, these constraints are robust to 21cm foreground removal, including the removal of systematic foreground wedge modes.

In addition to constraints on the standard cosmological parameters, the cross-bispectrum provides competitive constraints on extended model parameters beyond the $w_0 w_a$CDM model considered above, which we explore in more detail in \cite{naidoo2022}. Here, we vary the sum of neutrino masses in a flat $\Lambda$CDM model and a flat $w_0 w_a$CDM model. After marginalising over the other cosmological parameters, we find a 1$\sigma$ constraint on the neutrino mass sum of 27.8 meV (29.0 meV) in a flat $\Lambda$CDM ($w_0 w_a$CDM) model, which improves on recent neutrino mass constraints of 72 meV at 2$\sigma$ \cite{adame2025desi}, and is weaker than forecasts of 16 meV at 1$\sigma$ from a combined bispectrum and power spectrum study \cite{abazajian2015neutrino} and 11 meV at 1$\sigma$ from a multi-tracer power spectrum study \cite{ballardini2022constraining}, all within the $\Lambda$CDM model. The neutrino mass constraint is sensitive to various assumptions \cite{qu2025unified}, for example, the value of the fiducial matter density, with a larger assumed fiducial matter density constraining the neutrino mass more tightly \cite{loverde2024massive, lynch2025s}. To study the impact of this assumption, we reduced the fiducial value of the matter density to $0.29$ (the 1$\sigma$ limit allowed by Planck) and found that the constraint weakens to 37.1 meV (37.6 meV) in a flat $\Lambda$CDM ($w_0 w_a$CDM) model. These constraints will improve significantly with future CMB lensing surveys. This demonstrates the cosmological constraining power of the \honenosp-CMB lensing cross-bispectrum. 

{\it Acknowledgements:} We acknowledge useful discussions with Devin Crichton, Martin Bucher, Pedro Ferreira, David Alonso, Roy Maartens, Louis Perenon, Francisco Villaescusa-Navarro, David Spergel, Will Coulton, Oliver Philcox, Eleanor Rath, Rakshitha Thaman and Anthony Pullen. We acknowledge visitor and sabbatical support from the Simons Foundation, where part of this work was completed. KM acknowledges support from the National Research Foundation of South Africa and the Fulbright Scholar Program. WN acknowledges support from the South African Radio Astronomy Observatory. 

\section{\large Supplementary Material}

\section{APPENDIX A: Why the two-point correlation of \hone intensity and a CMB secondary signal is negligible}

We show here that the cross-correlation between the CMB lensing signal, which has a broad redshift kernel, and an \hone intensity map, which has been cleaned of foregrounds and thus lacking long-wavelength radial modes, is negligible in the flat-sky limit due to the lack of overlap in large-scale radial modes. 

\begin{figure}[t!]
    \centering
    \includegraphics[scale=0.1]{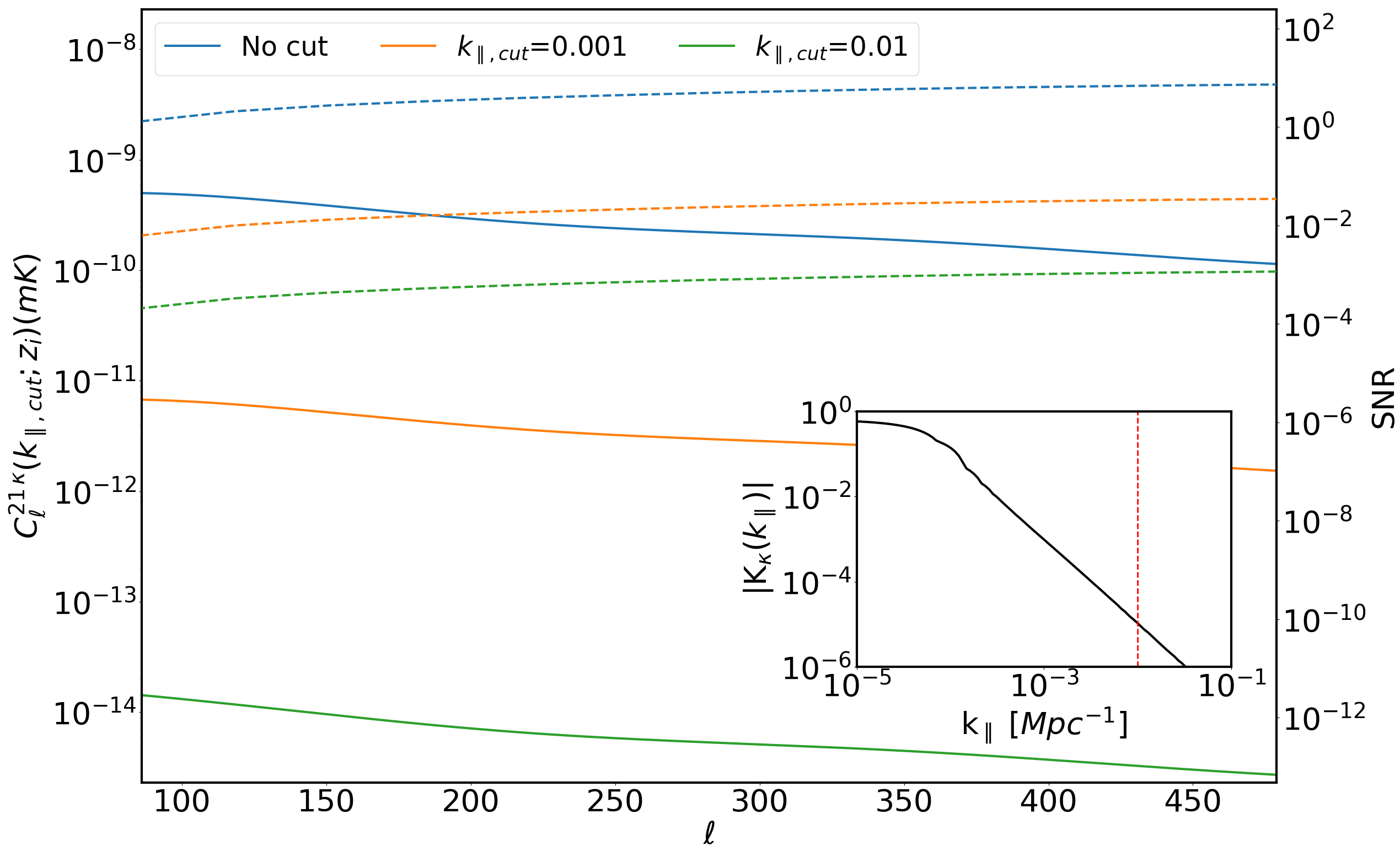}
    \caption{\hone-CMB lensing cross-correlation signal (solid lines) as a function of angular wavenumber for three different values of $k_{\parallel,cut}$ computed in the $z_i=0.95$ redshift bin. The corresponding dashed lines show the signal-to-noise ratio (SNR) for each case. In the inset plot, we show the CMB lensing kernel in Fourier space, which rapidly falls off with increasing $k_\parallel,$ and the nominal value of $k_{\parallel,cut}$ that we use in this letter. }
    \label{HI_kappa_sig_noise_zbin_kcut}
\end{figure}

The CMB lensing kernel has broad redshift support between today and last scattering, which means that most of the lensing signal is contained in low $\kpar \sim \chi_{*}^{-1}$ modes, as shown in the Figure \ref{HI_kappa_sig_noise_zbin_kcut} inset. Note that our description of the CMB lensing signal in terms of $\kpar$ modes is approximate, due to evolution within the broad redshift bin. However, ${\tilde K}_\kappa(\kpar),$ the radial transform of $K_\kappa(\chi),$ is only used here for conceptual value and a more precise treatment would require a light-cone decomposition into suitable angular and radial modes.

The lensing convergence and \hone IM cross-correlation spectrum in redshift bin $z_i$ is given by
\begin{equation}
C_{S,i}^{\kappa \dt21}(\ell,y) =
\bTb (z_i) \zoneHI(\bfk;z_i) K_\kappa
\left
({y \over \rnui }
\right
) 
{P_{m,0}
\left
( {\ell \over \cpli}, {y \over \rnui }
\right
) \over  V_p(z_i)/D(z_i)} 
  \nonumber
\end{equation}
where $P_{m,0}$ is the matter power spectrum at redshift zero and $K_\kappa(y/\rnui)$ is the real part (due to the power spectrum symmetry) of ${\tilde K}_\kappa$.
The key point here is that the lensing kernel is a function of $y$ modes probed by the \hone field, which do not include the lowest frequency modes ($\kpar \lesssim k_{\parallel,cut}$) removed in the foreground cleaning process. Indeed, in Figure \ref{HI_kappa_sig_noise_zbin_kcut} we see that for higher values of $k_{\parallel,cut}$ the cross-spectrum is significantly reduced due to the rapid fall-off in $K_\kappa$ with increasing $y.$ 

\section{APPENDIX B: Survey experiments, power spectra and signal-to-noise ratio}

In Appendix A, we saw that the cross-correlation spectrum between CMB lensing and \hone intensity is drastically reduced when low $\kpar$ \hone modes are removed in the foreground cleaning process. We can quantify this loss of signal in terms of the reduced cross-correlation signal-to-noise for the CMB and \hone IM experiments considered in this letter, specifically AdvACT \cite{Henderson_2016} 
and HIRAX \cite{newburgh2016hirax}, which will overlap over $\sim$ 15,000 deg$^2.$ 

HIRAX is a radio interferometer array of 6m dishes currently under construction in South Africa, which will measure the \hone intensity mapping signal in the 400-800 MHz band, while AdvACT was a 6m mm-wave telescope operating in Chile that made arcminute resolution maps of the CMB. The instrument and survey specifications for these experiments given in Table \ref{table:Expt_specs} are used to specify the power spectrum noise for each experiment. 

\begin{table}[!!!!!!!!!!!!h]
	\centering
	\begin{tabular}{|c|c|}
		\hline
		HIRAX \hone IM 
		& AdvACT CMB lensing \\
		\hline
		$S_{\rm area}=$ 15,000\,$\text{deg}^2$ 
		&  $S_{\rm area}=$ 15,000\,$\text{deg}^2$\\
		\hline
	    $t_\text{tot}=$4\,yrs; $f_{\rm eff}$ = 0.5
	    & Channels: 90, 150 GHz  \\
		\hline
		Bandwidth=0.4-0.8\,GHz
		&  Beam FWHM = 2.2, 1.4 arcmin \\
		\hline
         $T_{\text{sys}}=$ 50\,K 
		& $T_{\rm map} = 8, 7$ $\mu K$-arcmin \\
		\hline
		$N_{\text{dish}}=$ 1024; $D_{\text{dish}}=$6\,m
		& $P_{\rm map}$ = 11, 10 $\mu K$-arcmin   \\
		\hline
	\end{tabular}
	\caption{Experimental specifications for the AdvACT \cite{Henderson_2016} and HIRAX \cite{newburgh2016hirax} surveys considered in this letter.}
	\label{table:Expt_specs}
\end{table}

For the \hone IM survey the noise is given by \cite{bull2015late}
\begin{equation}
C_{N,i}^{\dt21}(\ell, y) =  {T_{\rm sys}^2(\tnu_i) \,{\rm S_{area}} \lambda^4 \over  \nu_{21} \,n_{\rm pol} \,t_{\rm obs} {\rm FOV}(\tnu_i) A_e^2 \, n\left({\bf u} = {\bfl / 2\pi}\right)},
\label{eq:interferom_noise1}
\nonumber
\end{equation}
where $T_{\rm sys}$ is the system temperature, ${\rm S_{area}}$ is the total survey area, ${\rm FOV} \approx \left({\lambda \over D_{dish}}\right)^2$ is the frequency-dependent field of view, $A_e$ is the dish collecting area, $n_{\rm pol}=2$ as HIRAX will have dual polarization feeds, $t_{\rm obs} = f_{\rm eff}\, t_{\rm tot}$ is the effective survey time, and $n({\bf u})$ is the baseline density in $uv$ coordinates.
The CMB lensing noise is given by \cite{hu2002mass}
\begin{equation}
\begin{aligned}
& C_{N}^{\kappa}(\ell)= 
\frac{\ell^4}{4} \left[\int \frac{d^2\ell'}{(2\pi)^2} \right. \times \\&  \left. \frac{[\bfl'.\bfl \, C_{S}^{EB}(\ell) + (\bfl-\bfl').\bfl \, C_{S}^{EB}(| \ell'-\ell |)]^2 \sin^2(2\phi) }{ {C}^{EB}_{tot} (\ell){C}^{EB}_{tot}(| \ell'-\ell |)}\right]^{-1}
\end{aligned}
\nonumber
\end{equation}
where ${C}^{EB}_{tot} (\ell) = C_{S}^{EB}(\ell)  + C_{N}^{EB}(\ell) $ and $\phi$ is the angle between $\bfl$ and $\bfl - \bfl'.$
We only consider the $EB$ estimator since it provides a close to optimal reconstruction \cite{okamoto2003cosmic} but will consider the full lensing constraining power of more sensitive CMB surveys in a follow-up paper \cite{naidoo2022}.

The signal-to-noise ratio for a given probe or cross-probe in redshift bin $z_i,$
\begin{flalign}
\hspace{-0.25cm} ({\rm SNR}_i)^2 = & {\Delta \tilde{\nu}_i  S_{\rm area}  \over 2} \int_{y_{\rm min}}^ {y_{\rm max}} {d y \over (2 \pi)}
\int_{\ell_{\rm min}}^ {\ell_{\rm max}}{\ell d \ell \over (2 \pi)} 
{\mathcal{S}_i(\ell, y)^2 \over \mathcal{V}_i(\ell,y)}, 
\label{eq:general_SNR}
\end{flalign}
is obtained by integrating over independent transverse and radial modes with relevant volume factors. Here, ${\mathcal{S}_i(\ell, y)}$ is the probe signal power spectrum and ${\mathcal{V}_i(\ell,y)}$ is the variance of that signal. The lensing convergence and \hone IM angular power spectra are given by 
\begin{eqnarray} 
\begin{aligned}
& C_{S,i}^{\dt21}(\ell, y) = \bTb^2 (z_i)\, \zoneHI^2 \, (\bfk; z_i){ P_m \left(\bfk, z_i\right)/ \,V_p(z_i)},\\
& C_{S}^{\kappa}(\ell) = \int d \cpl W_\kappa(\cpl)^2 {P_m(\bfk, \chi)\over \cpl^2},   \label{eq: HI Power Spectrum}
\nonumber
\end{aligned}
\end{eqnarray}
where the Limber approximation is used for the CMB lensing power spectrum expression. 

The signal-to-noise ratio in Equation \ref{eq:general_SNR} is specified as follows for the different power spectra. For the \hone power spectrum, the signal and variance terms are given by ${\mathcal{S}^{\dt21}_i(\ell, y)} = C_{S,i}^{\dt21}(\ell, y)$ and 
$\mathcal{V}^{\dt21}_i(\ell,y) =  \left[C_{S,i}^{\dt21}(\ell, y) + C_{N,i}^{\dt21}(\ell, y)\right]^2.$ The lensing convergence signal and variance terms are given by ${\mathcal{S}_i^{\kappa}(\ell)} = C_{S}^{\kappa}(\ell)$ and 
$\mathcal{V}_i^{\kappa}(\ell) =  \left[C_{S}^{\kappa}(\ell) + C_{N}^{\kappa}(\ell)\right]^2$ and we do not integrate over $y$ modes. For the \hone IM-CMB lensing cross-correlation, the integrand is given by
\begin{equation}
\begin{aligned}
{ {C_{S,i}^{\kappa \dt21}(\ell, y)}^2 \over  {C_{S,i}^{\kappa \dt21}(\ell, y)}^2 + \left[C_{S}^{\kappa}(\ell) + C_{N}^{\kappa}(\ell)\right]\left[C_{S,i}^{\dt21}(\ell, y) + C_{N,i}^{\dt21}(\ell, y)\right]}.\nn
\end{aligned}
\end{equation}
 
We restrict the $\ell$ and $y$ integration ranges used in this letter to the {\it linear} scales accessible by the HIRAX HI and AdvACT CMB lensing surveys, as described in more detail in \cite{naidoo2022} \footnote{Specifically, we find that over the HIRAX redshift range, $\ell_{min}$ varies from about 60 to 100, $y_{min}$ from about 90 to 120, $\ell_{max}$ from about 800 to 1100, and $y_{max}$ from about 1400 to 2000.}. 
In Figure \ref{HI_kappa_sig_noise_zbin_kcut} we show the cumulative cross-correlation SNR in a given redshift bin as a function of angular modes and integrated over all radial modes, for different foreground cuts. We note that for $k_{\parallel,cut}=0.01$ Mpc$^{-1}$ the cross-correlation SNR drops by several orders of magnitude, thus severely degrading its detectability.

\section{APPENDIX C: Derivation of the \honenosp-\honenosp-$\kappa$ cross-bispectrum in the squeezed limit}
The cross-bispectrum estimator presented in this paper relies on gravity-induced higher order correlations between the density field to recover the long-wavelength \hone modes \cite{bernardeau2002large}. The bispectrum of first-order Gaussian fields vanishes, so at second order, we have $
\dt21 (\bfk; z_i) = \dtone21 (\bfk; z_i) + {\dttwo21} (\bfk; z_i),
$
where the second order contribution is
\begin{equation}
\begin{aligned}
\dttwo21 (\bfk, \cpl) &=  \int{ d^3 q \over (2\pi)^3} \, {\ztwoHI (\bfq, \bfk - \bfq,\cpl)\over \bTb(z_i)} \times \\ & {\dtone21 (\bfk  - \bfq , \cpl)\over \zoneHI (\bfk-\bfq, \cpl)} \, {\dtone21 (\bfq , \cpl) \over \zoneHI (\bfq, \cpl)}. \nn
\end{aligned}
\end{equation}

The second order redshift-space \hone kernel is given by \cite{bernardeau2002large} 
\begin{eqnarray}
&& \ztwoHI (\bfq, \bfk - \bfq, \cpl) = {1\over 2}\btwoHI(\cpl) +   {1\over 2}  f(\cpl) \kpar \times \nn \\ && \left[ {\mu_1 \over q_1}\left(  \boneHI(\cpl) + f(\cpl)  \mu_2^2\right) 
	+  {\mu_2 \over q_2} \left(  \boneHI(\cpl) + f(\cpl)  \mu_1^2\right)\right] \nn \\
&& + \boneHI(\cpl) F_2(\bfq,\bfk-\bfq) + f(\cpl) \left( {\kpar \over k }\right)^2 G_2(\bfq,\bfk-\bfq) \nn 
\end{eqnarray}
with 
\begin{eqnarray}
& \mu_1 = {q_\parallel \over q_1}, \quad q_1 = |\bfq| ~ ; \quad \mu_2 = {k_\parallel - q_\parallel \over q_2}, \quad q_2 =  | \bfk - \bfq |, \nn \\
 & F_2(\bfk_1,\bfk_2) = {5\over 7} +{2\over 7}  {(\bfk_1 \cdot \bfk_2)^2 \over k_1^2 \, k_2^2} + {1\over 2} 
 {\bfk_1 \cdot \bfk_2 \over k_1 \, k_2} \left( {k_1 \over k_2} +  {k_2 \over k_1}\right), \nn \\
& G_2(\bfk_1,\bfk_2) = {3\over 7} +{4\over 7}  {(\bfk_1 \cdot \bfk_2)^2 \over k_1^2 \, k_2^2} + {1\over 2} 
 {\bfk_1 \cdot \bfk_2 \over k_1 \, k_2} \left( {k_1 \over k_2} +  {k_2 \over k_1}\right),\nn
\end{eqnarray}
where the $F_2$ kernel given in \cite{bernardeau2002large} has been corrected (see e.g., \cite{munshi2017integrated}). 
For the \hone bias we use the redshift dependent form for $\boneHI$ and $\btwoHI$ from \cite{penin2018scale}. In the above we ignore higher order non-linear corrections to $Z_{HI}^{(2)}$ \cite{bernardeau2002large,karagiannis2020forecasts}
as these effects are sub-dominant for the linear scales we consider here and in the squeezed limit these terms, including the tidal bias, vanish exactly \cite{naidoo2022}. As indicated above, in this study we restrict ourselves to linear \hone and CMB lensing modes that contribute to the bispectrum. 

We choose to correlate the power spectrum of the local \hone temperature field, the so-called position-dependent \hone power spectrum,
\begin{equation}
\begin{aligned}
\left. P_{21}(\bfk; z_i)  \right\vert_\bfr
	= {1\over V_{L}} 
\left. \dt21\left (\bfk; z_i \right)   \right\vert_\bfr
\left. \dt21^*\left (\bfk; z_i \right)  \right\vert_\bfr
\nonumber
\end{aligned}
\end{equation}
in some volume, $V_{L}(z_i)= L_\parallel L_\perp^2=\Omega_i \Delta \tilde \nu_i  V_p(z_i),$ centred at position $\bfr$ in redshift bin $z_i,$ with the mean density of that volume, as traced by the CMB lensing convergence field, $\kappa.$ Specifically, we correlate the {\it projected} \hone power spectrum, $P_{21}(\ell, y; z_i)  \vert_\bfr = {1\over V_{p}} P_{21}(\bfk; z_i)  \vert_\bfr,$ with the average lensing convergence in that volume.
As discussed in \cite{chiang2014position}, this correlation defines an integrated bispectrum. The position-dependent power spectrum probes coupling between large-scale and small-scale modes by measuring the local power spectrum, which is correlated with the mean density in that volume. Here the local \hone field in the volume (up to second order) is given by
\begin{equation}
\hspace{-0.25cm}
\dt21(\bfk; z_i) \vert_\bfr
=V_{L} \int {d^3 k_1 \over (2\pi)^3} e^{-i\bfk_1 \cdot \bfr} W_{L}^{21}(\bfk_1) \dt21\left (\bfk - \bfk_1; z_i \right)
\nonumber
\end{equation}
where we choose top-hat window functions in position space, corresponding to sinc functions in harmonic space, $W_{L}(q) = \sinc (q).$ The average CMB lensing convergence in the volume is obtained by transforming the lensing convergence, $\kappa (\bft; z_i),$ and taking the $\ell=0$ mode, which gives
\begin{equation}
\bar{\kappa}(z_i) \vert_\bfr = {V_{L}  W_\kappa(\cpli) \over  \cpli^2} \int {d^3 q' \over (2\pi)^3} e^{-i \bfr \cdot \bfq'}  W_{L}^{\kappa}(\bfq') \delta_m(-\bfq'; z_i).
\nonumber
\end{equation}
In practice, lensing maps contain the integrated lensing signal from last scattering to today, however, the cross-correlation of the lensing signal outside the \hone redshift bin will vanish, hence we can consider just the lensing signal within the bin. We are careful, though, to calculate the lensing variance from the full lensing signal integrated along the line of sight.

Using these expressions in the cross-bispectrum expression and simplifying, we obtain
\begin{equation}
\begin{aligned}
&B_{S,i}^{\bar \kappa \dt21 \dt21}(\ell, y; z_i) = \llangle  \left. P_{21}(\ell, y; z_i)  \right\vert_\bfr \bar{\kappa}(z_i) \vert_\bfr \rrangle \\
& = {V_L^2 W_\kappa(\cpli) D^4(\cpli)\over V_p \cpli^2} \int {d^3 q \over (2\pi)^3} W_{L,\kappa}(\mathbf{q}) P_m(\mathbf{q};z=0) \\ & \int {d^3 k_1  \over (2\pi)^3} W_{L,21}(\mathbf{k_1}) W_{L,21}^*(\mathbf{k_1} + \bf{q})    \times  \\
&  \left\{ P_{21}(\mathbf{k}'; z=0) {Z_{HI}^{(2)}(\mathbf{q}, \mathbf{k}') \over Z_{HI}^{(1)}(\mathbf{k}') }  +  P_{21}(\tilde{\mathbf{k}} ; z=0) {Z_{HI}^{(2)}(\mathbf{-q}, \tilde{\mathbf{k}}  ) \over Z_{HI}^{(1)}(\tilde{\mathbf{k}} ) }\right\}
\end{aligned}
\end{equation}
where we have set $\mathbf{k}' = \mathbf{k} - \mathbf{k_1} - \mathbf{q}$ and $\tilde{\mathbf{k}} = \mathbf{k} - \mathbf{k_1} $. 

We are interested in the squeezed limit of the cross-bispectrum estimator, which corresponds to $k \gg q$ and $k \gg k_1.$ In this limit, we obtain 
\begin{equation}
\begin{aligned}
B_{S,i}^{\bar\kappa \dt21 \dt21}(\ell, y) = V_B(\cpli) P_{21}(\mathbf{k};z_i) \mathcal{B}(k, \mu_k; f, b_{HI}^{(1)}, b_{HI}^{(2)}) \\ \times \int q^2 {dq} W_{L,\kappa}(q) W_{L,21}(q) P_m(q;z_i).
\end{aligned}
\end{equation}
where we have used
$\int {d^3 k_1 \over (2\pi)^3}  W_{L,21}(\mathbf{k_1}) W_{L,21}^*(\mathbf{k_1} + \mathbf{q})  = W_{L,21}(\mathbf{q})/V_L$ and assumed that the window functions are isotropic. In the above, we have defined $V_B(\cpli) = {V_L W_\kappa(\cpli) / (V_p \cpli^2)}$ and 
\begin{equation}
\begin{aligned}
	\mathcal{B}(k, \mu_k, f, b_{HI}^{(1)}, b_{HI}^{(2)}) =  {1\over 3} \left( 3 - {d \log{P_m} \over d \log{k}} \right)  \left(f\mu_k^2 - \mu_k^2 + 2  \right)
    \\ + {1 \over 14 Z_{HI} } \left(14b_{HI}^{(1)}f\mu_k^2 + {14\over 3}b_{HI}^{(1)} f  + {26\over 3} b_{HI}^{(1)} \mu_k^2 + {26 \over 3}b_{HI}^{(1)} \right. \\ \left.+ 28b_{HI}^{(2)} + 14f^2\mu_k^4 - 14f^2\mu_k^2 - 6f\mu_k^4 + {38\over 3}f\mu_k^2 \right)
\end{aligned}
\end{equation}

\section{APPENDIX D: \honenosp-\honenosp-$\kappa$ cross-bispectrum covariance and signal-to-noise ratio}

The \honenosp-\honenosp-$\kappa$ cross-bispectrum SNR is given by Equation \ref{eq:general_SNR}, where 
${\mathcal{S}_i(\ell, y)} = B_{S,i}^{\bar\kappa \dt21 \dt21}(\ell, y; z_i)$ and 
$\mathcal{V}_i(\ell,y)=2 \mathcal{V}^{\dt21}_i(\ell,y) \, \mathcal{V}_i^{\kappa}(\ell)$ is the the Gaussian contribution to the variance. 
We have found that the Gaussian contribution dominates the diagonal covariance, being much greater than the diagonal covariance term containing the HI-$\kappa$ two-point correlation, which we have shown is negligible, and much greater than the diagonal contributions from the non-Gaussian `BB' and `PT' covariance terms, which can be significant for squeezed bispectra \cite{barreira2019squeezed, biagetti2021covariance, floss2022primordial}.
These terms are smaller than the Gaussian covariance term on large angular scales due to the dominant Gaussian signal term, as can be seen in \cite{barreira2019squeezed}, and on small angular scales due to the dominant lensing reconstruction and HI noise \cite{naidoo2022}. 
Off-diagonal contributions from the `BB' and `PT' terms are also small relative to the diagonal Gaussian contribution, but could affect cosmological parameter correlations in a nontrivial way. We have perturbatively included the off-diagonal contribution from the `BB' term, which dominates over the `PT' term on the scales we are interested in, and found negligible changes to the parameter constraints presented in the next section. A more detailed study of the cross-bispectrum covariance is presented in \cite{naidoo2022}.

\bibliography{21cmblens_bibtex}
\onecolumngrid

\end{document}